\documentclass[preprint2]{aastex62}
\pdfoutput=1
\usepackage{mathtools}
\usepackage{subfigure}

\graphicspath{{./}{figures/}}

\shorttitle{Discovery of a supercluster candidate at $z \sim 1.1$}
\shortauthors{Narwal et al.}

\begin{document}

\title{Discovery of a supercluster candidate at $z \sim 1.1$}

\correspondingauthor{Tapish Narwal}
\email{ph1150833@iitd.ac.in}

\author{Tapish Narwal}
\affil{Indian Institute of Technology Delhi\\ Hauz Khas, New Delhi, India 110001}

\author{Tomotsugu Goto}
\affiliation{National Tsing Hua University\\ No. 101, Section 2, Kuang-Fu Road\\ Hsinchu, Taiwan 30013}

\author{Tetsuya Hashimoto}
\affiliation{National Tsing Hua University\\ No. 101, Section 2, Kuang-Fu Road\\ Hsinchu, Taiwan 30013}
\affiliation{Centre for Informatics and Computation in Astronomy (CICA), National Tsing Hua University, 101, Section 2. Kuang-Fu Road, Hsinchu, 30013, Taiwan (R.O.C.)}

\author{Seong Jin Kim}
\affiliation{National Tsing Hua University\\ No. 101, Section 2, Kuang-Fu Road\\ Hsinchu, Taiwan 30013}

\author{Chia-Ying Chiang}
\affiliation{National Tsing Hua University\\ No. 101, Section 2, Kuang-Fu Road\\ Hsinchu, Taiwan 30013}

\author{Yi-Han Wu}
\affiliation{National Tsing Hua University\\ No. 101, Section 2, Kuang-Fu Road\\ Hsinchu, Taiwan 30013}



\begin{abstract}
We report a promising candidate for a distant supercluster at z $\sim 1.1$ in the Dark
Energy Survey Science Verification data.
We examine smoothed semi-3D galaxy density maps in various photo-z slices. 
Among several overdense regions, in this work we report the most significant one as having a $3\sigma$ overdensity at a redshift of $\sim1.1$, over a $\sim160$ Mpc scale, much larger than the regular cluster scale (several Mpc).
The shape of the supercluster is not circular in the sky projection. Therefore, we regard the point of maximum overdensity as the center for quantitative measurements.
Two different estimates suggest the mass of the supercluster to be $1.37\substack{+1.31 \\ -0.79} \times 10^{17} M_{\odot}$, more than one order more massive than regular galaxy clusters.
Except for protosuperclusters found with emission-line galaxies, this could be the most distant supercluster to date defined by regular galaxies. A spectroscopic confirmation would make this a very interesting object for cosmology.
We discuss the possible implications of such a massive structure for $\Lambda$CDM cosmology.

\end{abstract}

\keywords{supercluster -- large-scale structure of the universe -- galaxies}


\section{Introduction}
On a cosmic scale, galaxies are distributed in a filamentous large-scale structure, and superclusters are the largest aggregations of galaxies and galaxy clusters in the cosmos. The formation of such large regions of density enhancement in the context of $\Lambda$CDM cosmology is of great interest, especially for structures that were present in the early universe.
Overdensities on the megaparsec scale can be used to constrain cosmological constants like the initial matter distribution, $\sigma_8$, etc. \citep{Sheth2011}. Aside from cosmology, superclusters also make for interesting targets for various studies, such as looking at the effect of environment on parameters like star-formation rates, etc.
With the advent of deeper wide-area sky surveys, like the Dark Energy Survey (DES, \cite{2018arXiv180103181A}), we can probe into the early universe and look for large-scale overdensities in the galaxy density field, which can correspond to galaxy clusters or superclusters. We make the distinction between galaxy clusters and superclusters on the basis of size, with the size of a typical supercluster being $>30 h^{-1}$ Mpc, which is much larger than the size of a galaxy cluster ($<10 h^{-1}$ Mpc). We use this criterion because it is compatible with most of the definitions previous researchers have used to define superclusters in their papers.\newline Superclusters comprise multiple galaxy clusters, which in general are not expected to be gravitationally bound. A number of previous works reported superclusters in the local universe \citep{Einasto2006,Liivamaegi2012,Lietzen2016,Bagchi2017}, while some of them discovered superclusters at higher redshifts up to $\sim 0.9$, \citep{Lubin2000,Swinbank2007,Gilbank2008,Kim2016}  \newline
We begin our paper by explaining how our data were selected to create a suitable sample for analysis (Sect.~\ref{sec:data}). Then we move onto further treatment of the data and the methods employed to search for the supercluster and explain how we calculate some of the properties of the supercluster (Sect.~\ref{sec:method}). Finally, we detail some areas for improvement in our analysis and discuss the importance of this supercluster candidate and spectroscopic confirmation and the scope for future scientific work (Sect.~\ref{sec:discussion}).\newline
For the analysis in our paper, we defined our cosmology using the following cosmological constants, $H_0 = $ 70 km s$^{-1}$ Mpc$^{-1}$, $\Omega_{\Lambda}=0.69$, $\Omega_c=0.27$, $\Omega_b = 0.045$, $A_S = 2.1e-9$, $n_s= 0.96$. Other cosmological constants were derived from these.
\section{Data}
\label{sec:data}
We used data from the Dark Energy Survey Science Verification 1 (DES-SVA1 hereafter) data release. We chose to base our work on this because of the extensive depth of the survey data, and the large area of the sky it covers. Although our data only extend across $\sim140$ deg$^2$, we believe our method of analysis can be extended to the full data release as well, when it becomes available.
Our data from DES-SVA1 are concatenated with photometric redshifts (photo-z), from Skynet, a neural network trained to predict photo-z, provided with the data release by the DES team. We chose SkyNet based on the performance analysis done in \cite{Bonnett2016} against other photo-z algorithms. Figure 1 shows the photo-z distribution.\newline
We selected the appropriate continuous region of the sky from DES-SVA1 SPT-E: $\sim$140 deg$^2$ (see Fig. \ref{fig:all_galaxies}), overlapping the eastern part of the South Pole Telescope footprint \citep{Benson2014}.
We selected galaxies from the DES catalog using their MODEST CLASS classifier, which the DES team found to have $\geq90\%$ efficiency and purity. 
Only the best objects were selected using the BADFLAG and FLAGS\textunderscore[G,R,I,Z] parameters provided in the DES-SVA1 data release. 
Furthermore, galaxies with large errors ($> 1$ mag in any of the four bands, \textit{griz}), in the observed magnitude are removed to ensure accurate \textit{k}-correction and photo-z.\newline

Since we use photo-z with errors of $\sim0.1$, corresponding to $\sim220 $h$^{-1}$ Mpc at $z = 0.6$, \citep{Sanchez2014}, we partition our data set into redshift slices of width 0.20 in the redshift space, i.e. a slice might be from $z = 0.50$ to $z = 0.70$ etc. In our analysis, we restricted the redshift range between 0.20 and 1.24, based on the galaxy counts (Figure 1). Redshift slices were generated in an overlapping manner at intervals of 0.03; for example, two consecutive slices could be 0.60 - 0.80 and 0.63 - 0.83, hence objects might be detected in multiple redshift slices. This compensates for the photo-z errors \citep{Sanchez2016}. To calculate absolute magnitudes (M), apparent magnitudes (m) must be corrected for luminosity distance and \textit{K}-correction as 
\begin{equation}
	M\textsubscript{Q} = m\textsubscript{R} - DM(z) - K\textsubscript{QR}(z)
    \label{kcorrect}
\end{equation}
We use \textit{K}-correct v4\_3 \citep{Blanton2007}, to perform k corrections, using $z=0$ as the reference.\newline
To ensure completeness, especially at higher redshifts, we select galaxies that are brighter than $M\textsubscript{r}(AB)= -17.8$. We also set an upper limit of $M\textsubscript{r}(AB) = -21$. This is a conservative choice in our redshift range, since the median $10\sigma$ limiting magnitude in the $r$ band is $23.8$. This enables us to create a volume-limited sample and avoid the Malmquist bias.
All distances are measured on a comoving scale.
\begin{figure}
	\includegraphics[width=\columnwidth]{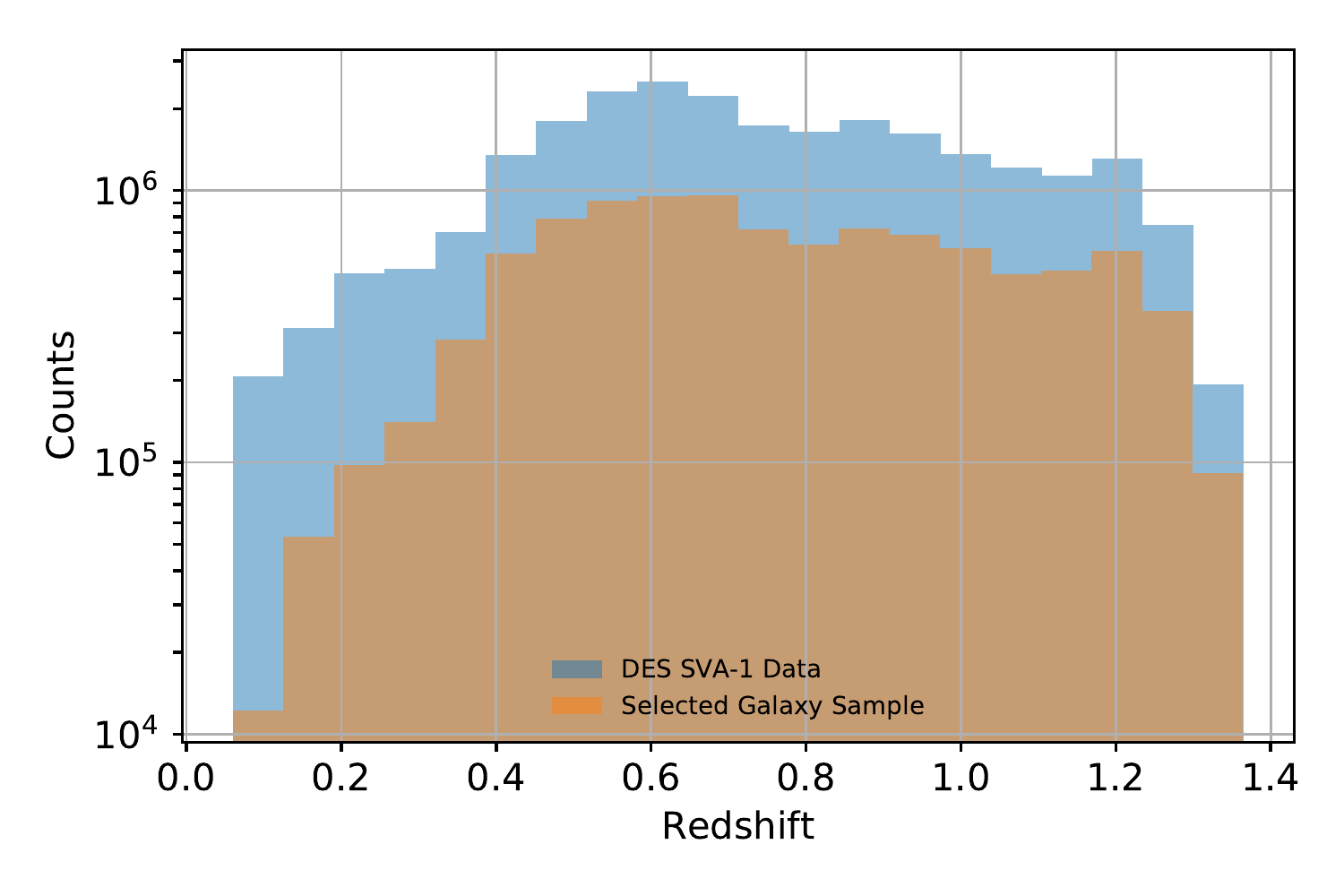}
    \caption{The redshift histogram of all the DES-SVA1 data (blue), and the selected galaxies (orange). It is noted that the observed counts start dropping quickly at $z>$1.2. }
    \label{fig:redshift_dist}
\end{figure}
\begin{figure}
	\includegraphics[width=\columnwidth]{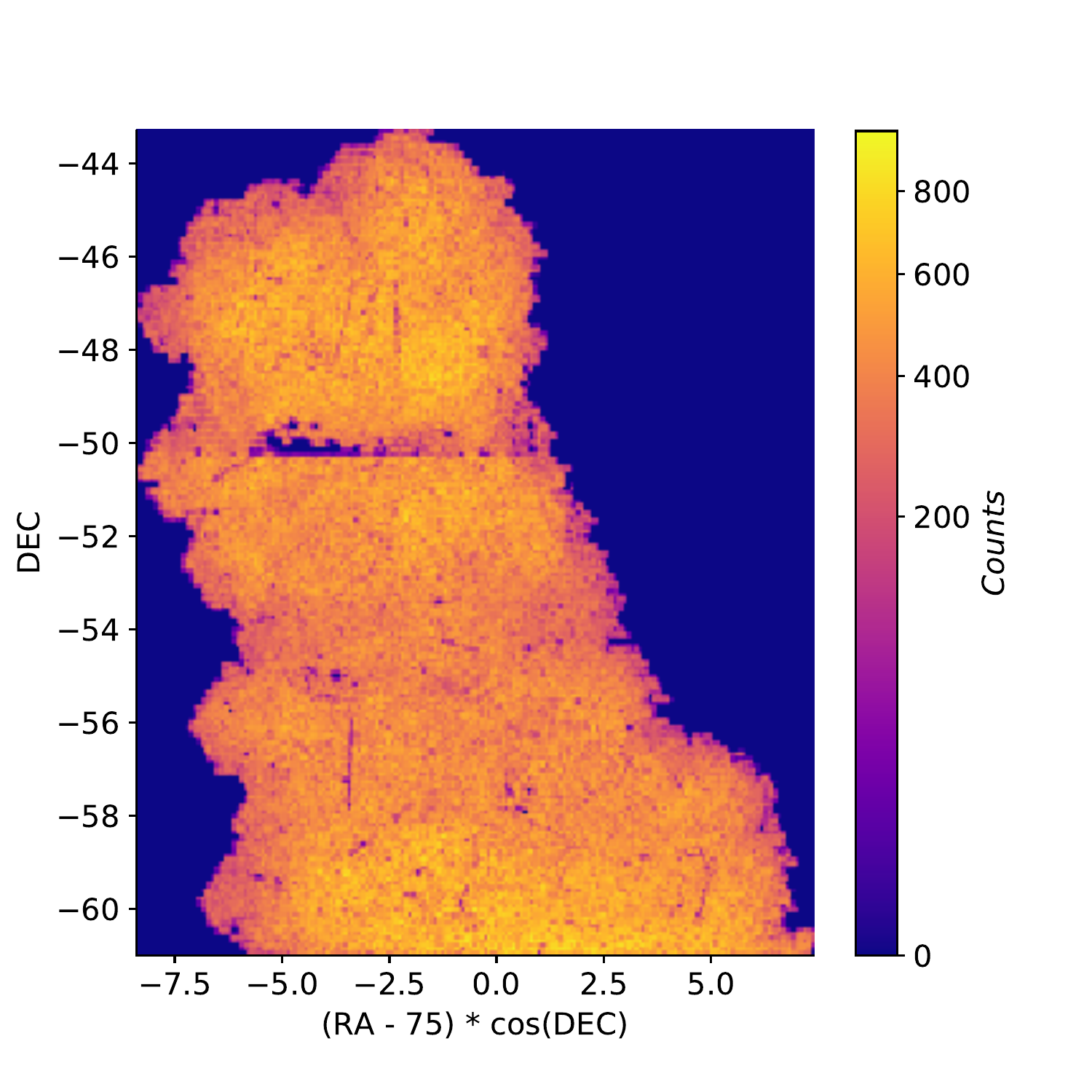}
    \caption{Galaxy counts map of the galaxies we have selected, across the full redshift range of the DES-SVA1.
    }
    \label{fig:all_galaxies}
\end{figure}
\section{Method}
\label{sec:method}
\subsection{Generating Galaxy Counts}
\begin{figure}
	\includegraphics[width=\columnwidth]{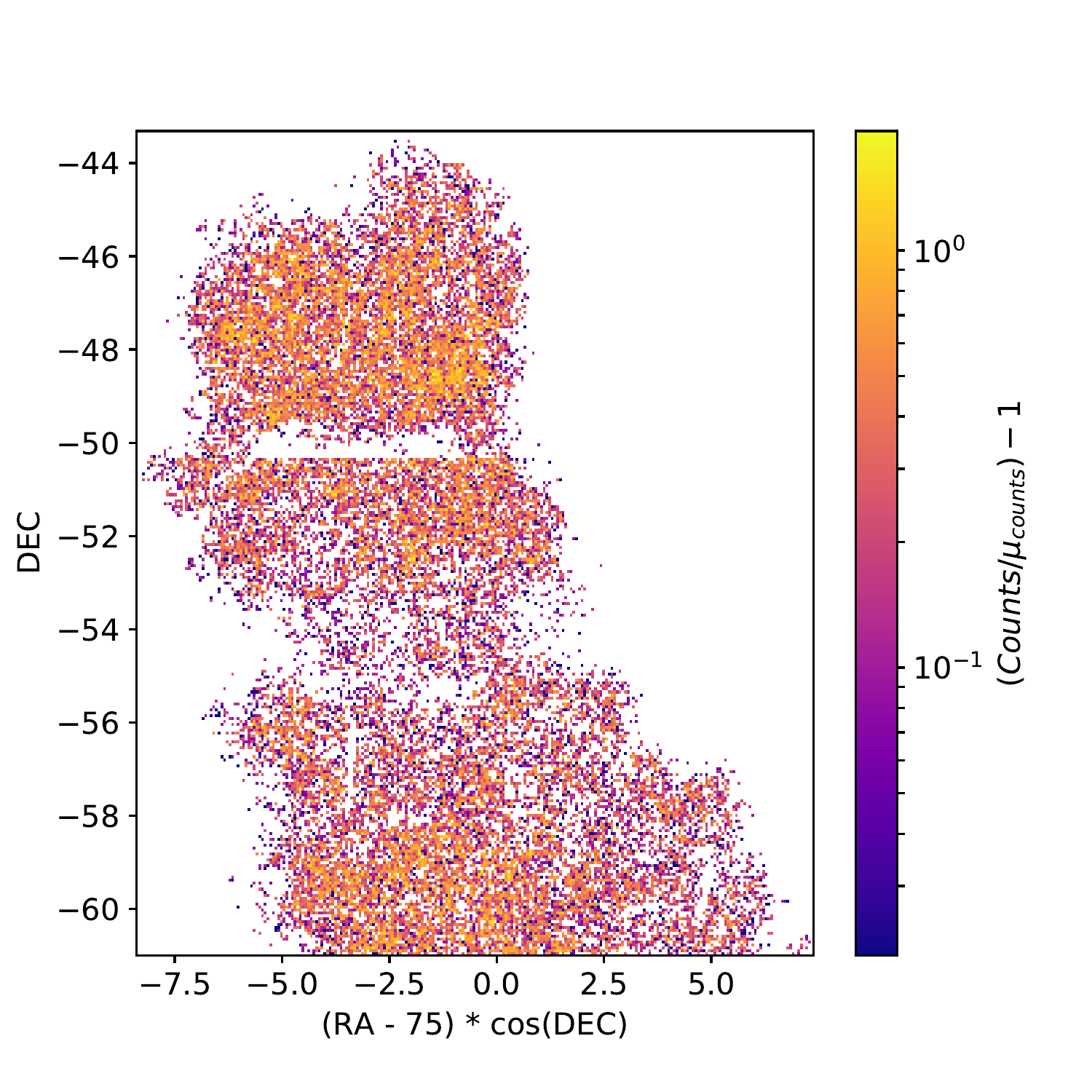}
    \caption{The ratio of galaxy counts in each pixel relative to the mean galaxy counts in $z=1.01-1.21$. We then smooth these counts to obtain fig. \ref{fig:contour}. In the \ref{sec:all_slices}ppendix other redshift ranges are presented after smoothing.}
    \label{fig:density_contrast}
\end{figure}

We binned galaxy positions in each redshift slice into square bins to generate a 2D histogram.
The bin size was fixed as 5 Mpc for our analysis. We also checked our result by varying the bin sizes and found that increasing the bin size to more than half the radial size of the supercluster causes severe underestimation of the overdensity because we lose contrast on the bins at the edge of supercluster regions. Smaller bin sizes have a negligible effect.
We count the number of galaxies in the volume-limited sample in each bin to generate a density contrast map as $\delta_c = (b_{c} / b_{\mu}) -1$, where $\delta_c$ is the density contrast, $b_{c}$ is the number of galaxies in a bin and $b_{\mu}$ is the average number of galaxies across all bins in a redshift slice.
The density contrast map is shown in Fig. \ref{fig:density_contrast}. We repeated this analysis in each redshift bin, and results are presented in the \ref{sec:all_slices}ppendix.
\subsection{Mask}
\begin{figure}
	\includegraphics[width=\columnwidth]{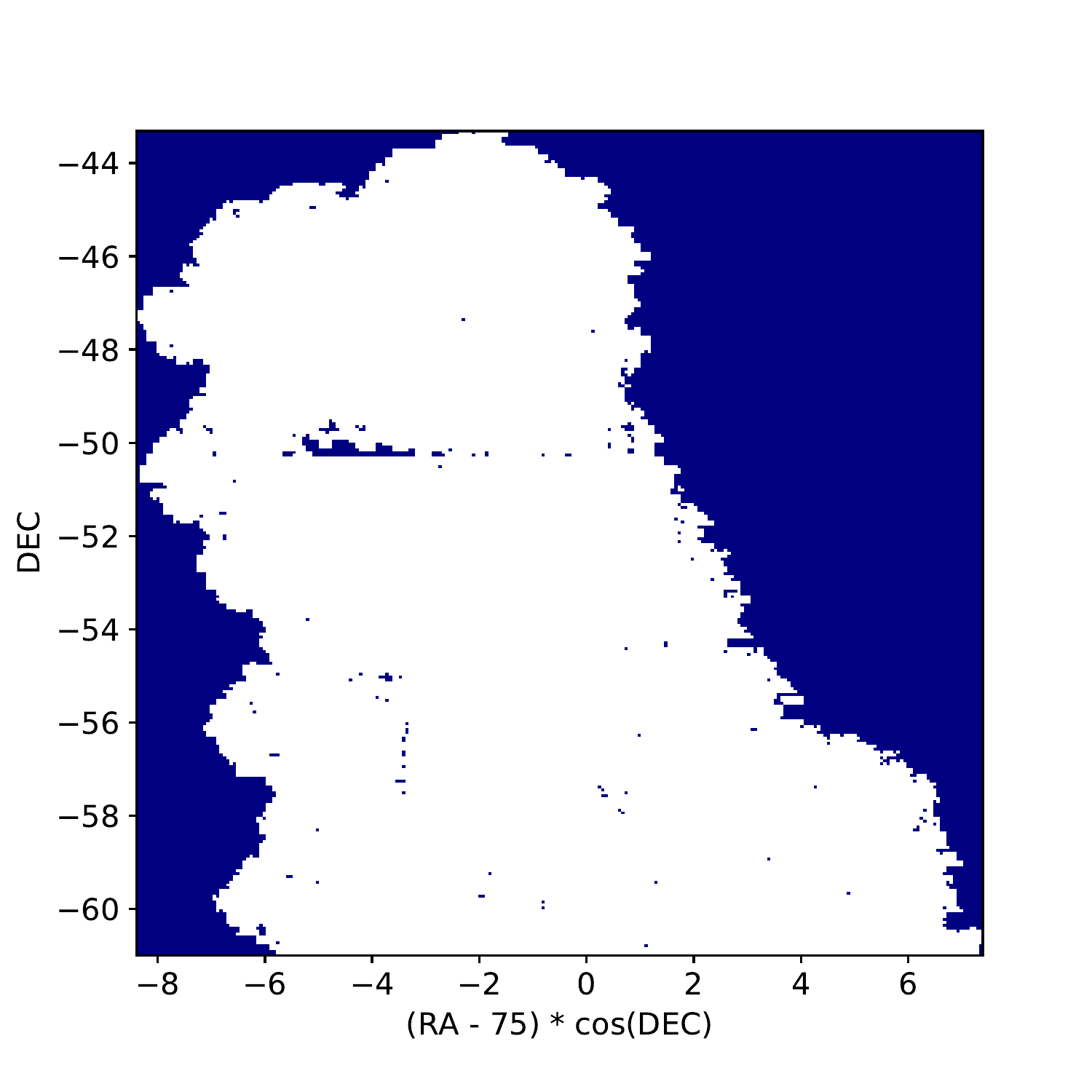}
    \caption{This figure shows the mask we have used. This is to prevent bins with low counts from affecting our analysis. The data in the blue region is the data that we have excluded.}
    \label{fig:mask}
\end{figure}

For each redshift bin that we analyze we generate a mask to help our analysis (Fig. \ref{fig:mask}). Our mask varies across redshift bins because the angular distance corresponding to a fixed comoving scale also differs with the redshift, and our analysis is done with a fixed comoving scale of $\sim5$ Mpc. \newline
Regions of the sky contaminated by bright stars or galaxies were already marked by the DES team in their catalog using using a column called BADFLAG. We selected only the cleanest parts of the sky for our analysis.
We generate our mask by looking at the galaxy count distribution in each 5 Mpc bin in our region of interest. We observe a Gaussian-like curve;however, there is a large number of bins where there have been no galaxies observed. We mask all bins with galaxy counts equal to zero. This provides us with our mask.
There are two main reasons for using a mask in our study:
\begin{enumerate}
\item Confinement-The region we have binned is much larger than the actual area covered in the DES-SVA1 SPT-E region. The SPT-E region has a nonuniform shape; hence when binning there is a need to mask regions that lack observational data.
\item Convolution-Since we later smooth our density contrast map, using convolution with a 2D Gaussian kernel, it is important that we mask bins without data, otherwise our convolution result will be underestimated due to the low counts in the unmasked regions. We discuss more about how to correct for the underestimation in the next section.
\end{enumerate}
\subsection{Smoothing}
\begin{figure*}
    \centering
	\includegraphics[width=\textwidth]{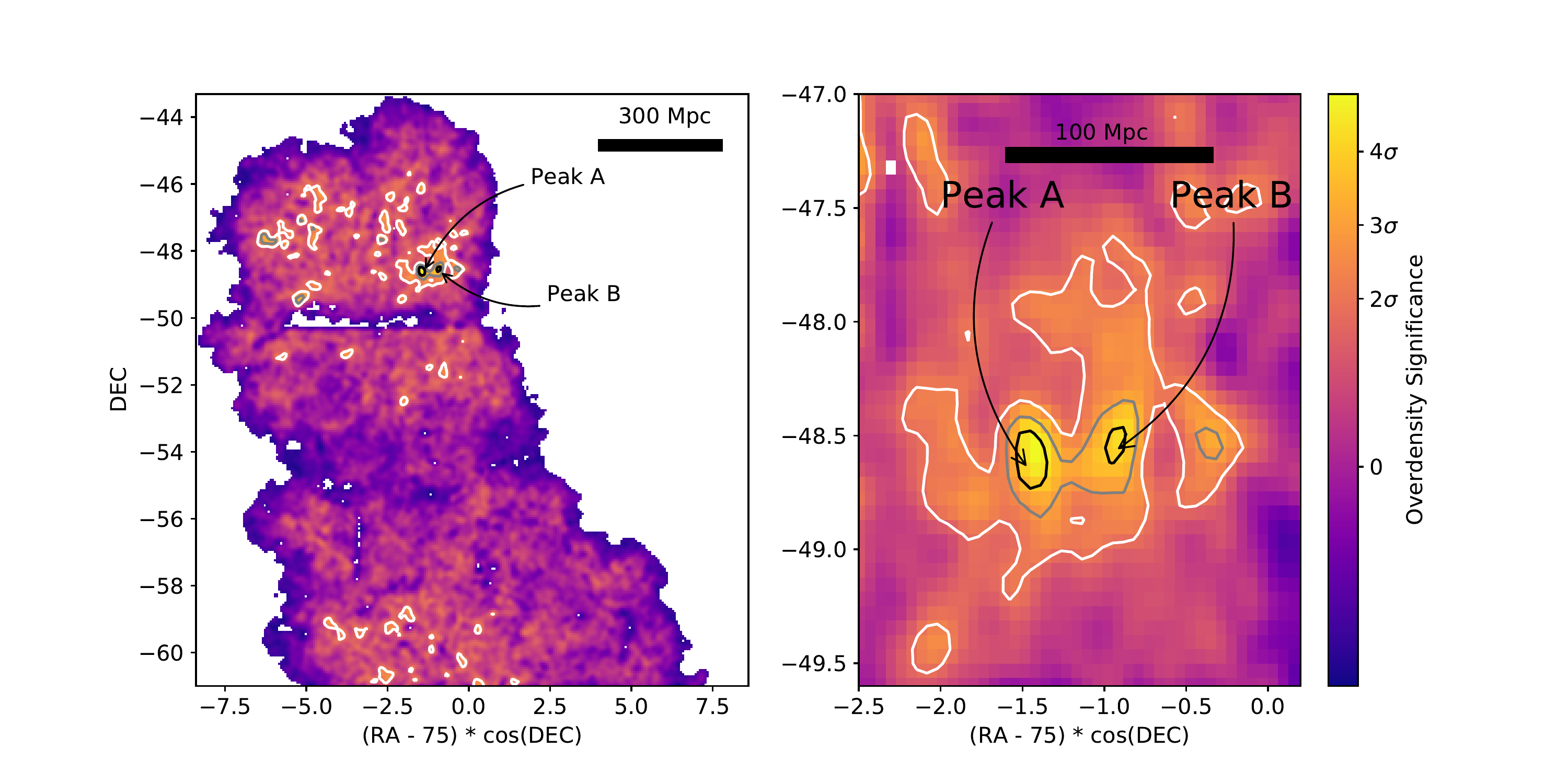}
    \caption{Contour plot of the density contrast smoothed with a Gaussian kernel, showing the significance of overdense structures in the redshift slice $1.04-1.24$, and a zoomed-in view of the supercluster region showing two distinct $4\sigma$ peaks. The scales of 300 Mpc and 100 Mpc are plotted for reference.}
    \label{fig:contour}
\end{figure*}

We used a 2D Gaussian kernel for our convolution, which is noted for good smoothing properties. Other choices of kernels could have been a Mexican hat kernel or a $B_3$ spline kernel. These kernels make sense for detection of galaxy clusters, due to the expectation of underdensities around galaxy clusters, but we refrain from using these because of the large scales we were working at, as compared to usual cluster size. \newline 
The mask helps with the convolution procedure, since the convolution treats the masked pixels as NaNs (not a number), and replaces those pixels with a kernel-weighted interpolation from neighboring pixels. This avoids underestimation of density due to regions where we have no data.
The size of the 2D Gaussian kernel needs to be of a scale similar to the structure we are looking for \citep{Einasto2014}. In our case we set the standard deviation of the Gaussian kernel as $6 Mpc$, however, previous studies show that results are not too sensitive to the smoothing scale \citep{Einasto2014}. Our calculations are consistent with this, showing low variation in the 4-8 Mpc range. However, we do note a significant loss of contrast at smoothing scales larger than 10 Mpc. This is perhaps due to excessive smoothing near the supercluster edges, which is exaggerated by the irregular shape of our candidate structure, and due to the loss of all internal structure of the superclusters, such as Peak A and B in Fig. \ref{fig:contour}.

\subsection{Finding the supercluster}
\begin{figure}
	\includegraphics[width=\columnwidth]{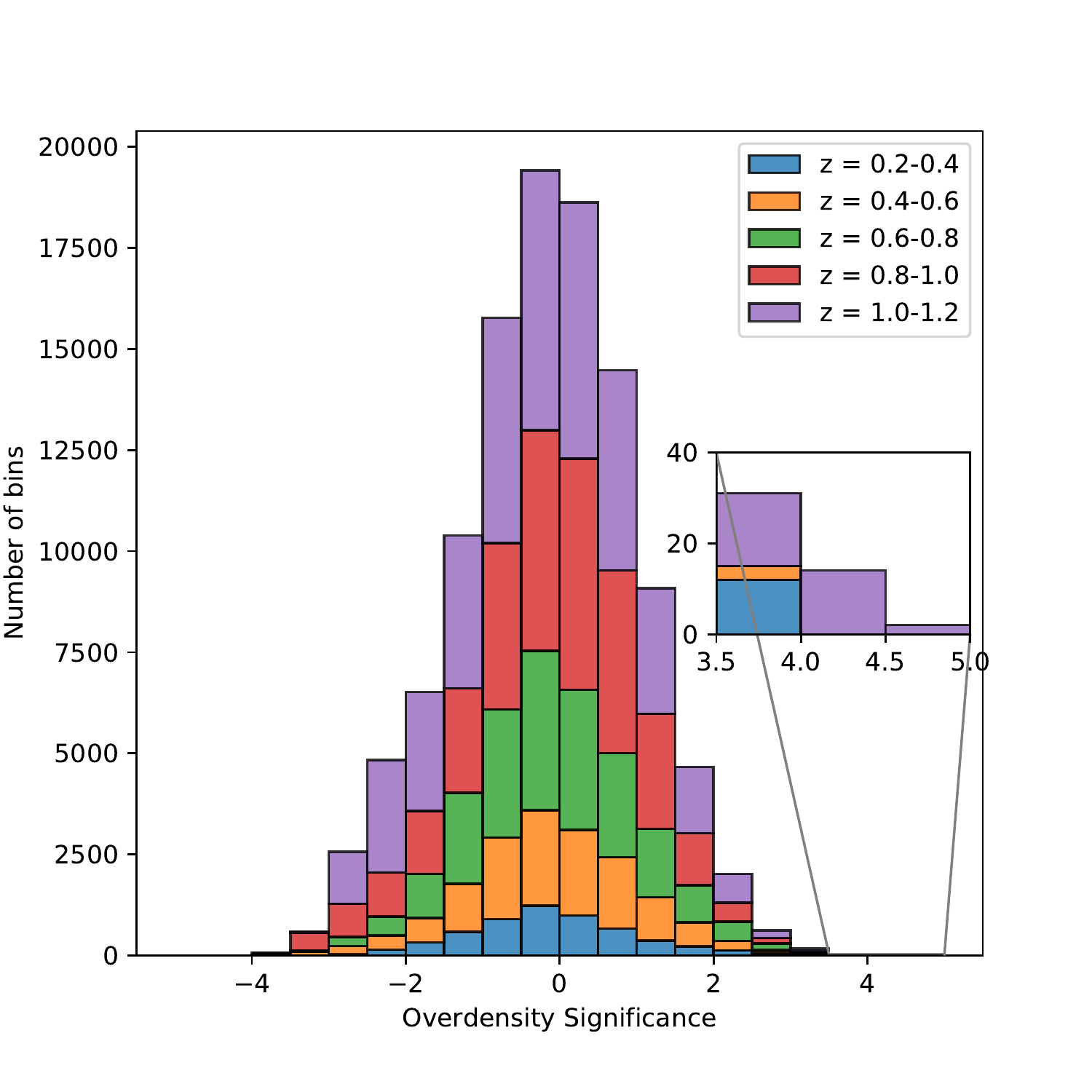}
    \caption{
    Overdensity significance distributions at each redshift bin. Bin size is fixed at 5 Mpc for all analyses.}
    \label{fig:overdensity_hist}
\end{figure}
We obtained overdensity contour plots for the galaxy distribution in all our redshift slices (Appendix \ref{sec:all_slices}). We then analyzed these structures across the redshift ranges and isolated supercluster candidates based on estimated size and significance.
We looked at the distribution of overdense pixels in different redshift ranges (Fig. \ref{fig:overdensity_hist}).
 We found that the structure in the redshift slice $z=1.04 - 1.24$ is the most overdense in the survey volume, with regions of more than $4 \sigma$ overdensity.
In this paper, we decided to focus and further investigate this structure.
Please see other less significant yet interesting supercluster candidates in the \ref{sec:all_slices}ppendix.

\subsection{Radius and Significance Estimation}
	\subsubsection{Radial Profile}
\begin{figure}
	\includegraphics[width=\columnwidth]{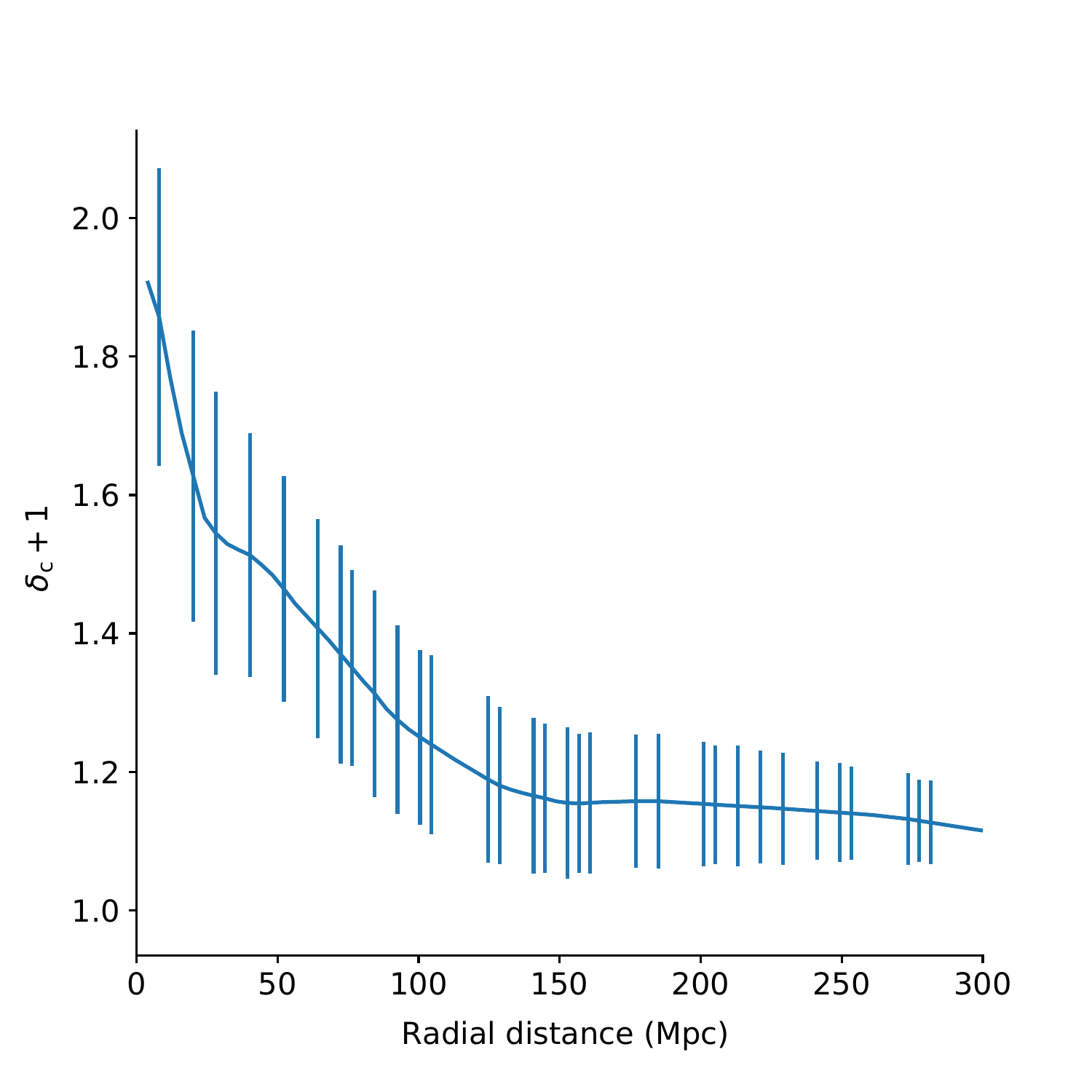}
    \caption{Radial profile of the supercluster region in a bin of $1.04<z<1.24$. Note that the shape is not quite circular, like that shown in Fig. \ref{fig:contour}.}
    \label{fig:radial_profile}
\end{figure}

        Finding/defining the center of our supercluster region is not straightforward. Here, we simply calculate the radial profile of the supercluster, in the redshift bin of $z = 1.04 - 1.24$, regarding the center of the most overdense pixel as the center. The radial profile is generated by calculating an azimuthal average over annular radial bins. 
        We show the derived radial profile in Fig. \ref{fig:radial_profile}.
        The slight bump at around 50 Mpc is attributed to the second cluster (peak B in Fig. \ref{fig:contour}), which is $\sim50$ Mpc away from the peak A.
       We note the break in the radial profile at $r \sim 150$ Mpc and that this radial profile does not follow a Navarro–Frenk–White (NFW) profile. \newline
       The density contrast in Fig. \ref{fig:radial_profile} is generated using the mean value of the density distribution in the survey area in a redshift slice. Similarly we use the standard deviation of the density distribution to make the error bars.
	\subsubsection{Significance Estimates}
\begin{figure}
	\includegraphics[width=\columnwidth]{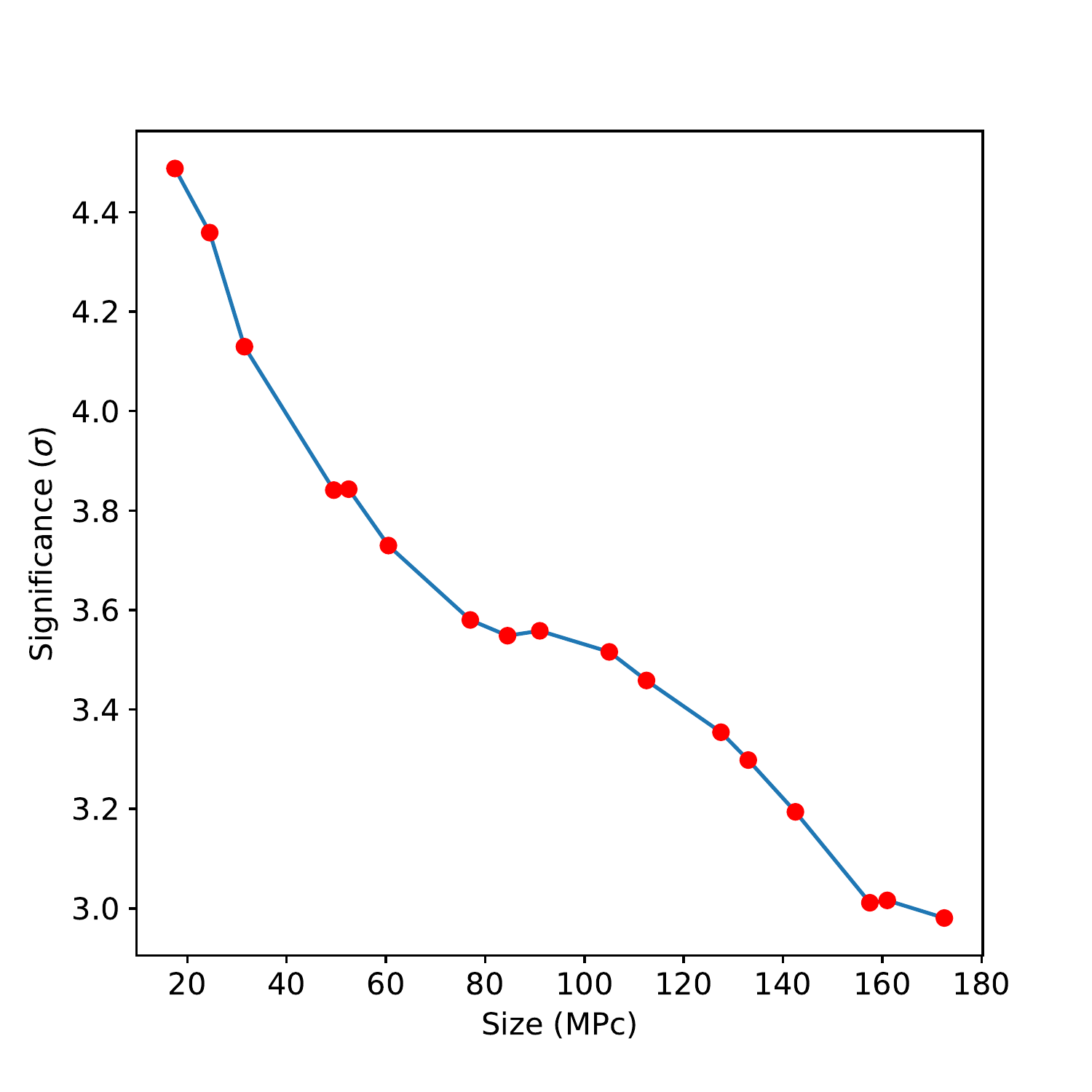}
    \caption{Significance of the overdensity as a function of the supercluster diameter in a bin of $1.04<z<1.24$.}
    \label{fig:significance_size}
\end{figure}

	To estimate the significance of our supercluster candidate, we analyze circular patches in the redshift bin of $z = 1.04 - 1.24$. For a patch of fixed radius $r$, we take the sum of all the pixels inside such a patch ($\Sigma_{pix}$). We look at the distribution of $\Sigma_{pix}$ of randomly placed patches of the same $r$, and compare the standard deviation of this distribution to the $\Sigma_{pix}$ of the supercluster region. This gives us the significance of the overdensity of our structure at radius $r$. \newline In case the patch contains masked pixels, we normalize $\Sigma_{pix}$ with respect to the number of unmasked pixels to ensure a meaningful analysis. Patches with more than half of their pixels masked are not used, to ensure minimal edge effects.\newline
	We repeat this analysis with patches of varying radius $r$ and obtain Fig. \ref{fig:significance_size}, which shows that the significance of the structure decreases with increasing radius. The decreasing nature in Fig. \ref{fig:significance_size} is consistent with what we roughly expect since it is evident from the radial profile in Fig. \ref{fig:radial_profile} that the density contrast decreases as we move away from the point of maximum overdensity. Hence it makes sense that the significance of the structure we are looking at also decreases as a function of its defined diameter. Also, the supercluster region is not circular in the 2-D projection, hence a circular patch includes low density regions which are outside the actual structure, so larger radii will cause increasing underestimation of the significance.\newline
	We can also observe that the profile becomes flatterer at around 160 Mpc (Figs.\ref{fig:radial_profile} and \ref{fig:significance_size}), which can be regarded as a measure of the size of the structure.
\begin{figure}
	\includegraphics[width=\columnwidth]{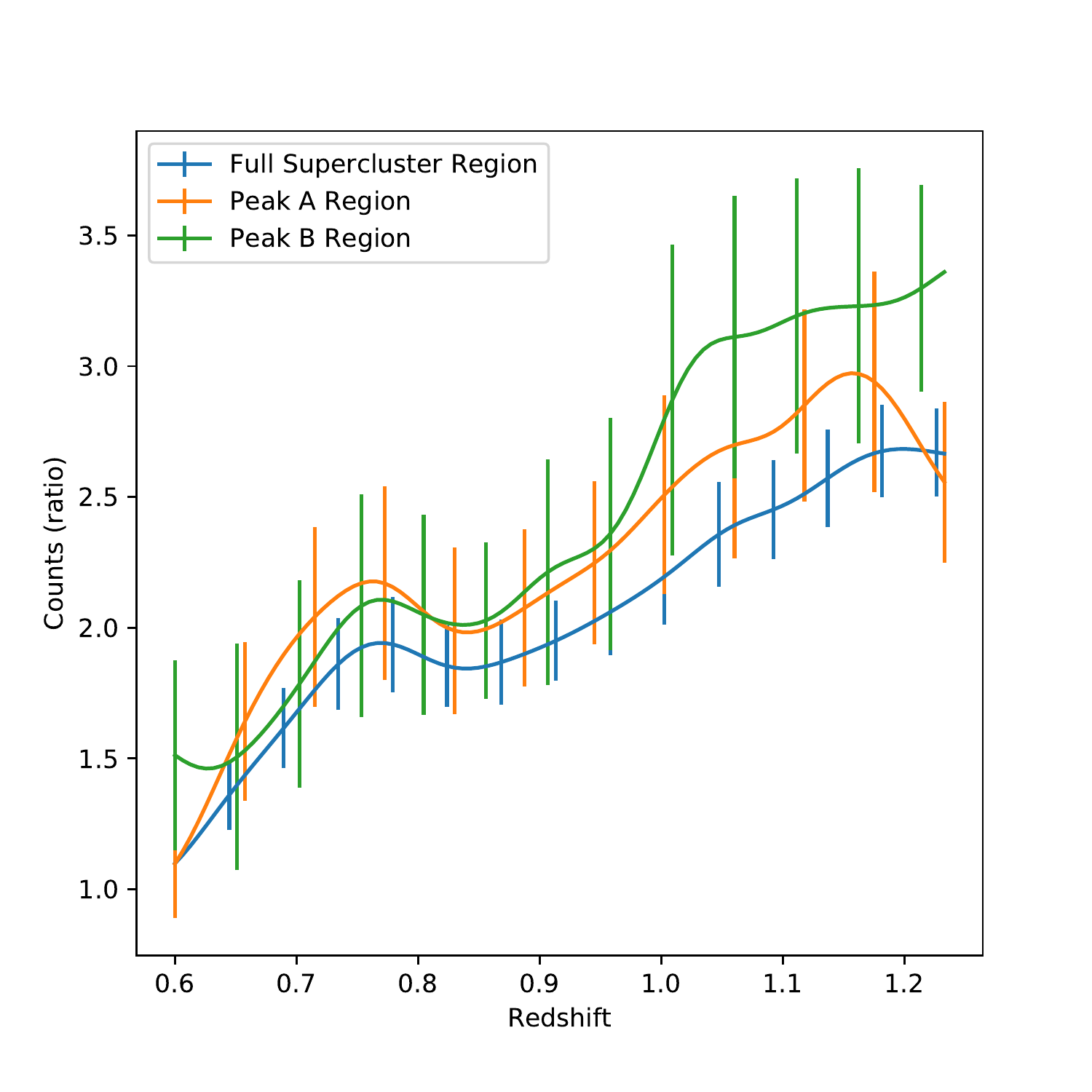}
    \caption{Relative galaxy counts with respect to field galaxies in the supercluster, Peak A, and Peak B regions as a function of redshift ($N_{region}/N_{field}$). Poisson error bars are included.}
    \label{fig:loc}
\end{figure}
\subsection{Estimating the redshift}
We look at relative galaxy counts in a patch near the most overdense region of the supercluster and the average counts in patches in the whole SPT-E region as functions of redshift. We select the size of the patch by observation of Fig. \ref{fig:contour} and the patch is defined as $72.5<RA<73.4$ and $-48.9<Dec<-48.2$. Fig. \ref{fig:loc} shows us a peak at redshift $\sim 1.15$, which can be regarded as a measure of the supercluster redshift.
There is also a peak at $z \sim 0.7$, which we found is due to another galaxy cluster in the same region of the sky as our supercluster candidate, along the line of sight. This peak corresponds to the overdensity in the same region of the sky as the supercluster candidate at a lower redshift, as can be seen in Fig. \ref{fig:7191}. \newline
Note that Fig. \ref{fig:loc} is restricted to redshift $<1.24$ since observed galaxy counts start decreasing rapidly after that (Fig. \ref{fig:redshift_dist}).
\subsection{Twin Peaks}
To further investigate our structure, we look at the relative galaxy counts in Peaks A and B as a function of redshift. We define and select the region of Peak A and Peak B as the $4\sigma$ overdense regions in Fig. \ref{fig:contour}.\newline
In Fig. \ref{fig:contour}, we see that there exists an overdensity in between Peaks A and B. The separation between A and B is $\sim35$ Mpc in RA-Dec space, which is seven bins in our analysis. The positional accuracy of the DES data is much better than our bin size of 5 Mpc. The significance of overdensity of these bins in between peak A and B is $\sim3 \sigma$. Therefore, we interpret these two structures as physically connected, suggesting that the existence of a supercluster.\newline
We see in Fig. \ref{fig:loc} that the redshift distributions of Peaks A and B are consistent with being from the same structure. It is not that one peak is at $z=1.04$, and the other is at $z=1.24$. Both peaks have similarly wide redshift distributions centered around $z=1.15$. This is also consistent with the interpretation that peaks A and B are physically connected. In Fig. \ref{fig:contour} we observe a large $2 \sigma$ overdense region surrounding Peaks A and B in RA-Dec space.\newline
From Fig. \ref{fig:loc}, the maxima at 1.05 is not a sharp peak, but rather a start of the increase of the density in redshift space, so we interpret it as an indicator of the advent of the supercluster region in redshift space, and not as one of peak A or B based on the evidence provided by the individual redshift distributions of peaks A and B, which are peaked around z $\sim1.15$.\newline

\subsection{Mass estimate}
We adopt two simple approaches to calculate the halo mass of our candidate supercluster since we have no membership probabilities to calculate mass richness from. For both of these, we need an idea of the diameter(size) of the supercluster. We make a rough estimate of the supercluster diameter by using Fig. \ref{fig:significance_size}. We note that a $3 \sigma$ overdensity corresponds to a $\sim160$ Mpc diameter for the presented supercluster candidate. We use this value for further calculations involving the size of the supercluster. For the error bars, we assume a $25\%$ error in our size estimate.
\begin{enumerate}
    \item Scaling Shapley supercluster mass-We assume the presented supercluster candidate to be spherical and take the halo mass calculated for the Shapely supercluster and scale it to the size of the presented supercluster candidate.
    Using the mass for the Shapley supercluster from \cite{Filippis2005}, we obtained a mass estimate of $1.10\substack{+1.03 \\ -0.63}  \times10^{17} M_{\odot}$ for the presented supercluster candidate. It is important to note that this method does not account for redshift evolution, and is probably an overestimate, as low-redshift superclusters are more collapsed and massive.
    \item Estimate from critical density-We use \cite{Steidel1998} to calculate the supercluster mass as follows:
    \begin{equation}
        M_{tot} = \rho_{m}\text{V}(1+\delta_{m})
    \end{equation}
    Here $\rho_{m}$ is the mean density of the universe.
    To obtain $\delta_m$ from $\delta_{<galaxy>}$, we use the $\delta_m = \delta_{<galaxy>} / b$ , where $b$ is the bias factor. We use bias factor $b = 1.1$ from \cite{Zehavi2011}. We use $\rho_m$ from (LSST DESC 2020, in preparation)( https://github.com/LSSTDESC/CCL). We hence obtain a supercluster mass estimate of
    $1.37\substack{+1.31 \\ -0.79} \times 10^{17} M_{\odot}$, which is consistent with that derived from the method (i).
\end{enumerate} 
\section{Discussion and Conclusions}
\label{sec:discussion}
\begin{figure}
	\includegraphics[width=\columnwidth]{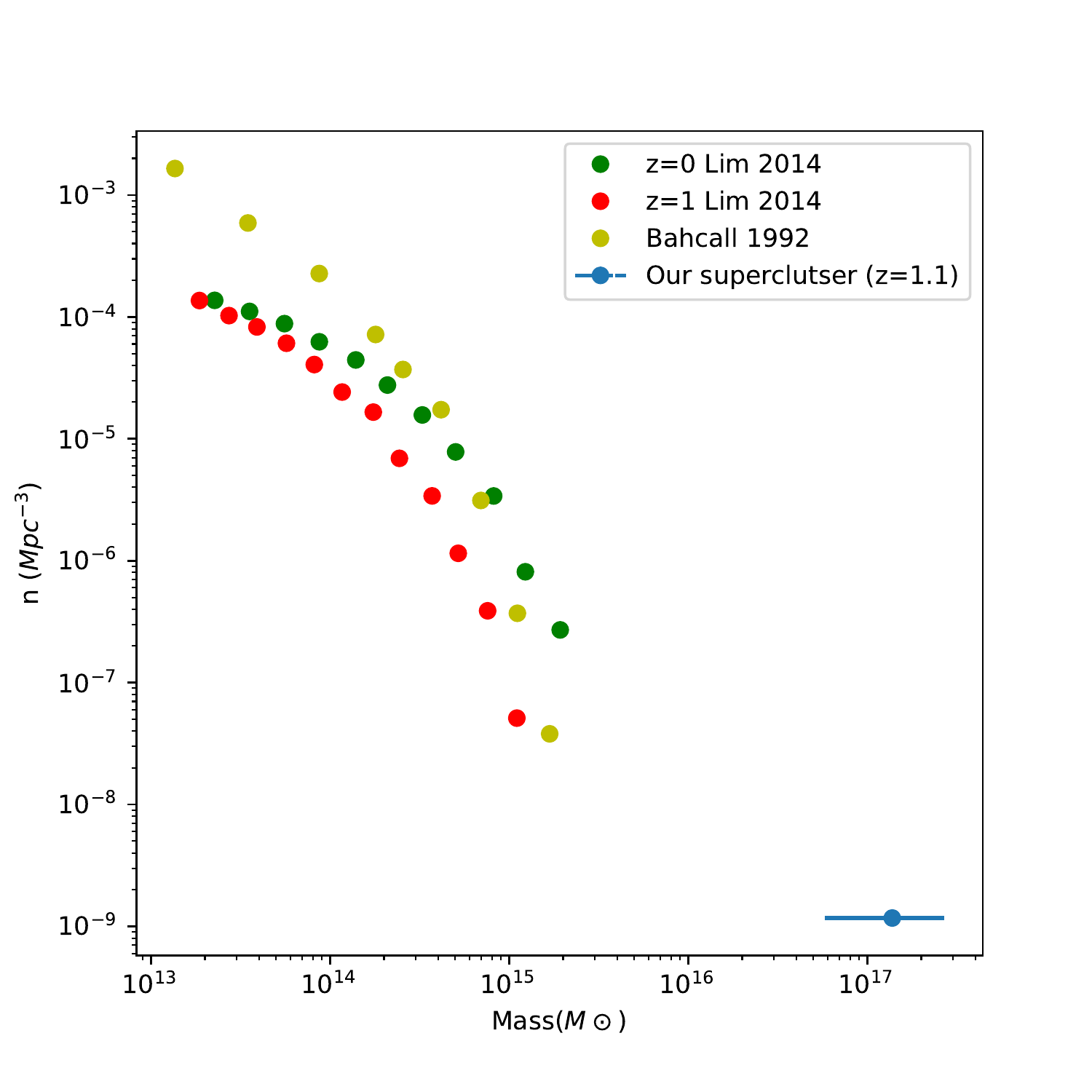}
    \caption{Comparison with supercluster mass functions in the literature, showing the number of superclusters in the survey volume as a function of the mass of the supercluster.}
    \label{fig:mass_func}
\end{figure}

We now discuss a few important factors in our study that affect the significance of our findings. 
While analyzing the regions of overdensity in a particular redshift slice, we do not calculate the counts of the field galaxies in the background of the supercluster in the same region of the sky, leading to a slight overestimation of supercluster membership. On the other hand when we generate circular patches to estimate the significance, we are underestimating the density of the structure, since it is not circular in the projection. 
Also, although using redshift slices mitigates most of the error due to the errors in the photo-z, it is not perfect and improved photo-z will help with further localization of the supercluster in redshift space. 
\newline
In our analysis we observe two regions of $>4\sigma$ overdensity (labeled Peak A and Peak B in Fig. \ref{fig:contour}) and both of them lie in the supercluster region in the redshift slice from 1.01 to 1.21. We note the relative galaxy counts in Peaks A and B as a function of redshift and compare them with the distribution for the whole supercluster region (Fig. \ref{fig:loc}). We find that both the peaks have a wide distribution in redshift space peaked around $z=1.15$. In Fig. \ref{fig:contour}, we can see that there exists an overdensity in between Peaks A and B. Therefore, we interpret these two structures as physically connected in RA-Dec space as well. This leads us to believe they are a part of the same larger structure. It seems like our candidate supercluster comprises of two large galaxy clusters with a filamentous region around them.\newline
A review of the literature shows us a wide variety in supercluster masses and sizes. \cite{Bagchi2017} find a massive ($10^{16} M_{\odot}$) supercluster at redshift $0.28$, with a size of $\sim200$ Mpc. \cite{Einasto2006} compiles more than 500 local superclusters at redshift $<0.2$ with effective diameters in the $20-50$ Mpc range. Researchers have tried to look for superclusters in multiple ways. \cite{Lubin2000} looked at a large population of red galaxies in the space separating two galaxy clusters at redshift $0.9$ and concluded the existence of a supercluster. \cite{Swinbank2007} used color selection to isolate overdensities across $30$ Mpc in the sky. \cite{Gilbank2008} optically selected overdensities and obtained mass estimates from X-ray temperatures of member clusters, finding a total mass of $\sim1.5\times10^{15} M_{\odot}$. \cite{Lietzen2016} identified an extended structure at redshift of $0.47$ with total mass of $2\times10^{17} M_{\odot}$. However, the total survey volumes searched are missing from most of the literature, so these are not shown in Fig. \ref{fig:mass_func}.\newline
Aside from protosuperclusters, defined by emission-line galaxies, this finding is a candidate to be the most distant supercluster to date, and as such is a powerful tool for probing into the early universe. It can be used as a good target for studies of the effects of environment on galaxies, etc.  \citep[e.g.,][]{2003MNRAS.346..601G}. \newline
We can also ask the question, whether we expect an overdensity like this in our cosmological models, since the most massive superclusters are thought to form from peaks in the initial mass-density fluctuations. It is interesting to ask whether or not this structure is virialized or where this structure lies in the evolutionary scale, toward local superclusters like the Shapley supercluster, or near protosuperclusters like the one mentioned in \cite{2018arXiv180606073C}. 
Confirmation of such a massive structure, at a high redshift of 1.1 using spectroscopic data, would be of great importance in the context of $\Lambda$CDM cosmology, especially if its existence is unexpected \citep{Sheth2011}. We would be more confident of the place of a supercluster like this in $\Lambda$CDM cosmology if we had reached higher redshifts so that could allow us to analyze the rarity of such objects.\newline

As evident from the superclusters in the literature, superclusters come in a huge range of sizes and masses, and we conclude that the size ($\diameter = 160$ Mpc) and mass estimate ($1.37\times10^{17}M_{\odot}$) of the presented supercluster candidate are not out of the ordinary. \cite{Sheth2011} discussed that the existence of Shapley supercluster ($\sim 10^{15} M_{\odot}$) is consistent with $\Lambda$CDM, and discussed that it is not unreasonable to find a structure as massive as ours in a much larger volume of the universe as compared to the local volume in which Shapley is enclosed. This, however, does not take into account cosmological evolution and we require simulations to estimate if such structures could have formed in the early universe corresponding to this candidate structure's redshift.\newline
Using data from \cite{Lim2014} for a supercluster mass function from simulations, and from \cite{Bahcall1992} for a galaxy cluster mass function, we compare the consistency of our object with $\Lambda$CDM. Simulations from \cite{Lim2014} also show the evolution of the supercluster mass function with redshift. Fig. \ref{fig:mass_func} compares these mass functions against our finding. Considering the mass of the presented supercluster candidate is one order larger than that in simulations, and how the redshift evolution of the supercluster mass function is a decreasing function of redshift \citep{Lim2014}, the existence of this supercluster might be inconsistent with $\Lambda$CDM cosmology. 
However, we must note that our mass estimate is highly inaccurate and the tension with the other mass functions might be reduced as we obtain an accurate mass estimate.
The number density of superclusters as massive as $10^{17} M_{\odot}$ is also quite uncertain statistically since we found only one such supercluster candidate at redshift ~1.1 in a volume of $~0.84\times10^{9}$ Mpc$^{3}$, which was our whole survey volume. A much wider survey volume could significantly affect our mass function plot.
Therefore, further deep and wider surveys are critically important to examine the $\Lambda$CDM cosmology in a more accurate way. Considering such wide-ranging implications on cosmology we think this is a prime candidate for spectroscopic confirmation.

\acknowledgments
\section*{Acknowledgements}
We thank the anonymous referee for careful reading of the paper, and many insightful comments.
T.G. acknowledges the support by the Ministry of Science and Technology of Taiwan through grant 105-2112-M-007-003-MY3 .\newline
T.N. acknowledges the support by the Ministry of Education of Taiwan through Taiwan Experience Education Program (TEEP).\newline
TH is supported by the Centre for Informatics and Computation in Astronomy (CICA) at National Tsing Hua University (NTHU) through a grant from the Ministry of Education of the Republic of China (Taiwan).\newline
This project used public archival data from the Dark Energy Survey (DES). Funding for the DES Projects has been provided by the U.S. Department of Energy, the U.S. National Science Foundation, the Ministry of Science and Education of Spain, the Science and Technology Facilities Council of the United Kingdom, the Higher Education Funding Council for England, the National Center for Supercomputing Applications at the University of Illinois at Urbana-Champaign, the Kavli Institute of Cosmological Physics at the University of Chicago, the Center for Cosmology and Astro-Particle Physics at the Ohio State University, the Mitchell Institute for Fundamental Physics and Astronomy at Texas A\&M University, Financiadora de Estudos e Projetos, Funda{\c c}{\~a}o Carlos Chagas Filho de Amparo {\`a} Pesquisa do Estado do Rio de Janeiro, Conselho Nacional de Desenvolvimento Cient{\'i}fico e Tecnol{\'o}gico and the Minist{\'e}rio da Ci{\^e}ncia, Tecnologia e Inova{\c c}{\~a}o, the Deutsche Forschungsgemeinschaft, and the Collaborating Institutions in the Dark Energy Survey.
The Collaborating Institutions are Argonne National Laboratory, the University of California at Santa Cruz, the University of Cambridge, Centro de Investigaciones Energ{\'e}ticas, Medioambientales y Tecnol{\'o}gicas-Madrid, the University of Chicago, University College London, the DES-Brazil Consortium, the University of Edinburgh, the Eidgen{\"o}ssische Technische Hochschule (ETH) Z{\"u}rich,  Fermi National Accelerator Laboratory, the University of Illinois at Urbana-Champaign, the Institut de Ci{\`e}ncies de l'Espai (IEEC/CSIC), the Institut de F{\'i}sica d'Altes Energies, Lawrence Berkeley National Laboratory, the Ludwig-Maximilians Universit{\"a}t M{\"u}nchen and the associated Excellence Cluster Universe, the University of Michigan, the National Optical Astronomy Observatory, the University of Nottingham, The Ohio State University, the OzDES Membership Consortium, the University of Pennsylvania, the University of Portsmouth, SLAC National Accelerator Laboratory, Stanford University, the University of Sussex, and Texas A\&M University.
This work is based in part on observations at Cerro Tololo Inter-American Observatory, National Optical Astronomy Observatory, which is operated by the Association of Universities for Research in Astronomy (AURA) under a cooperative agreement with the National Science Foundation.



\bibliographystyle{aasjournal}
\bibliography{ref.bib}



\appendix

\section{Other redshift slices}
\label{sec:all_slices}
\begin{figure*}%
\centering
\subfigure[][$z = 0.20 - 0.40$]{%
\label{fig:2040}%
\includegraphics[width=2in]{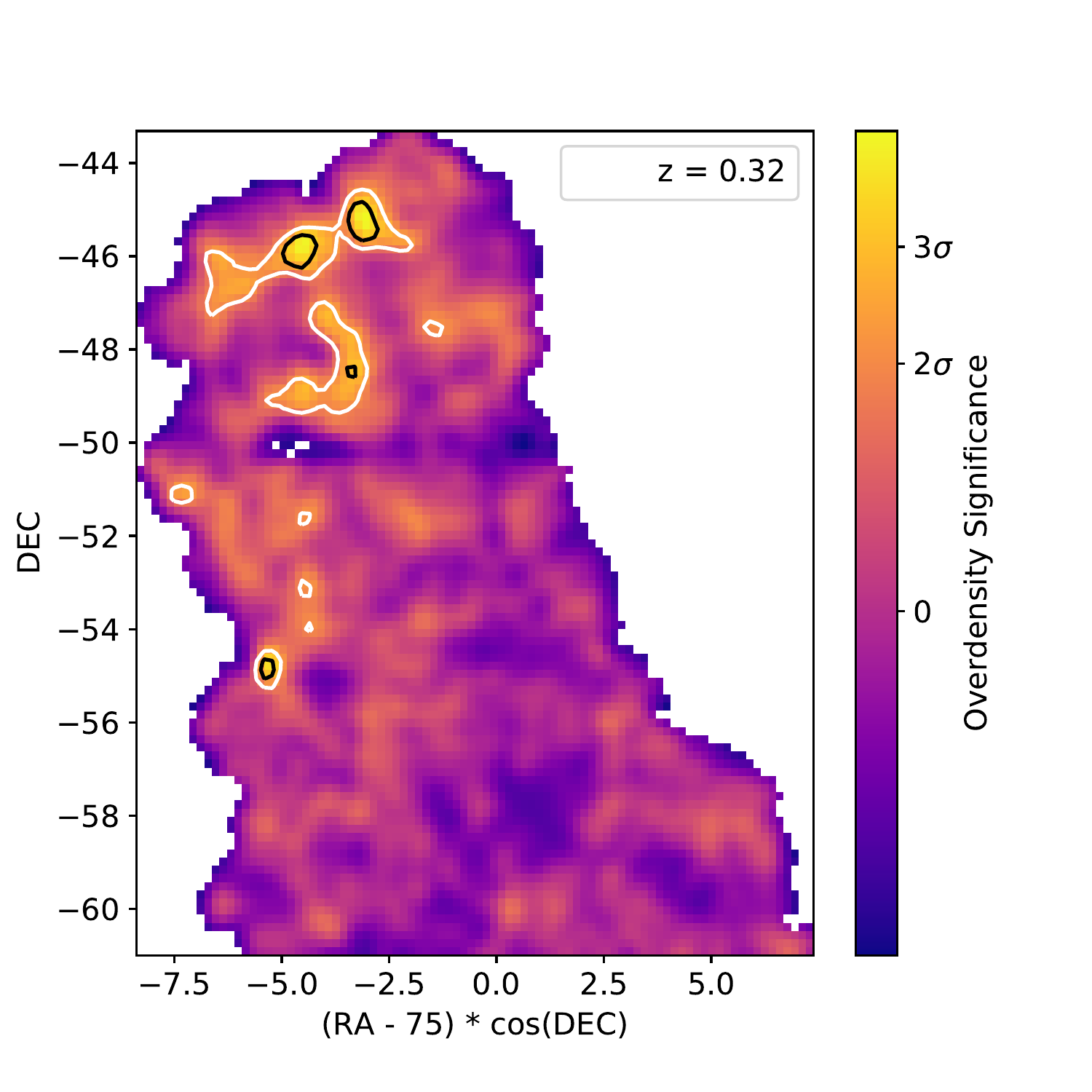}}%
\hspace{8pt}%
\subfigure[][$z = 0.29 - 0.49$]{%
\label{fig:2949}%
\includegraphics[width=2in]{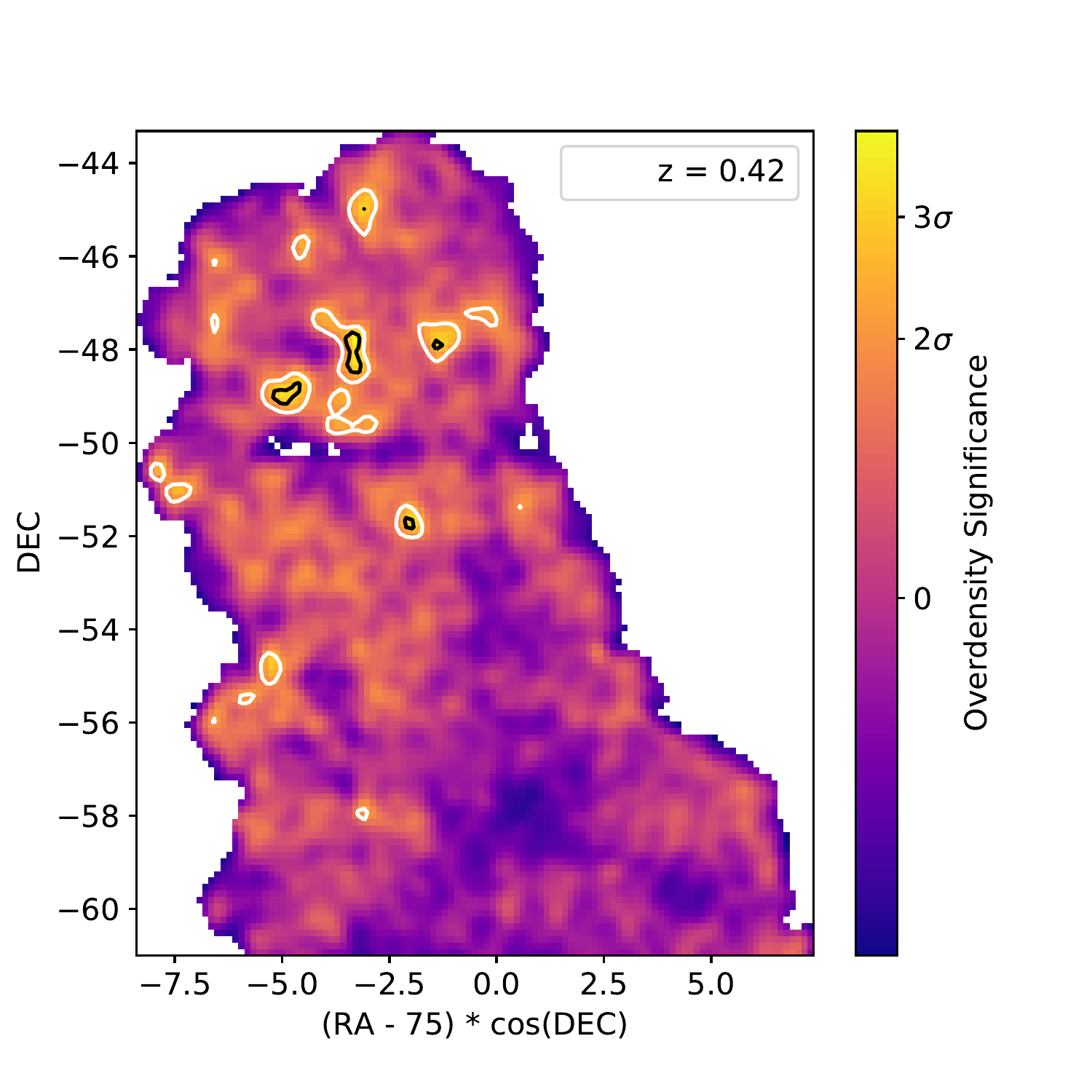}}%
\hspace{8pt}%
\subfigure[][$z = 0.41 - 0.61$]{%
\label{fig:4161}%
\includegraphics[width=2in]{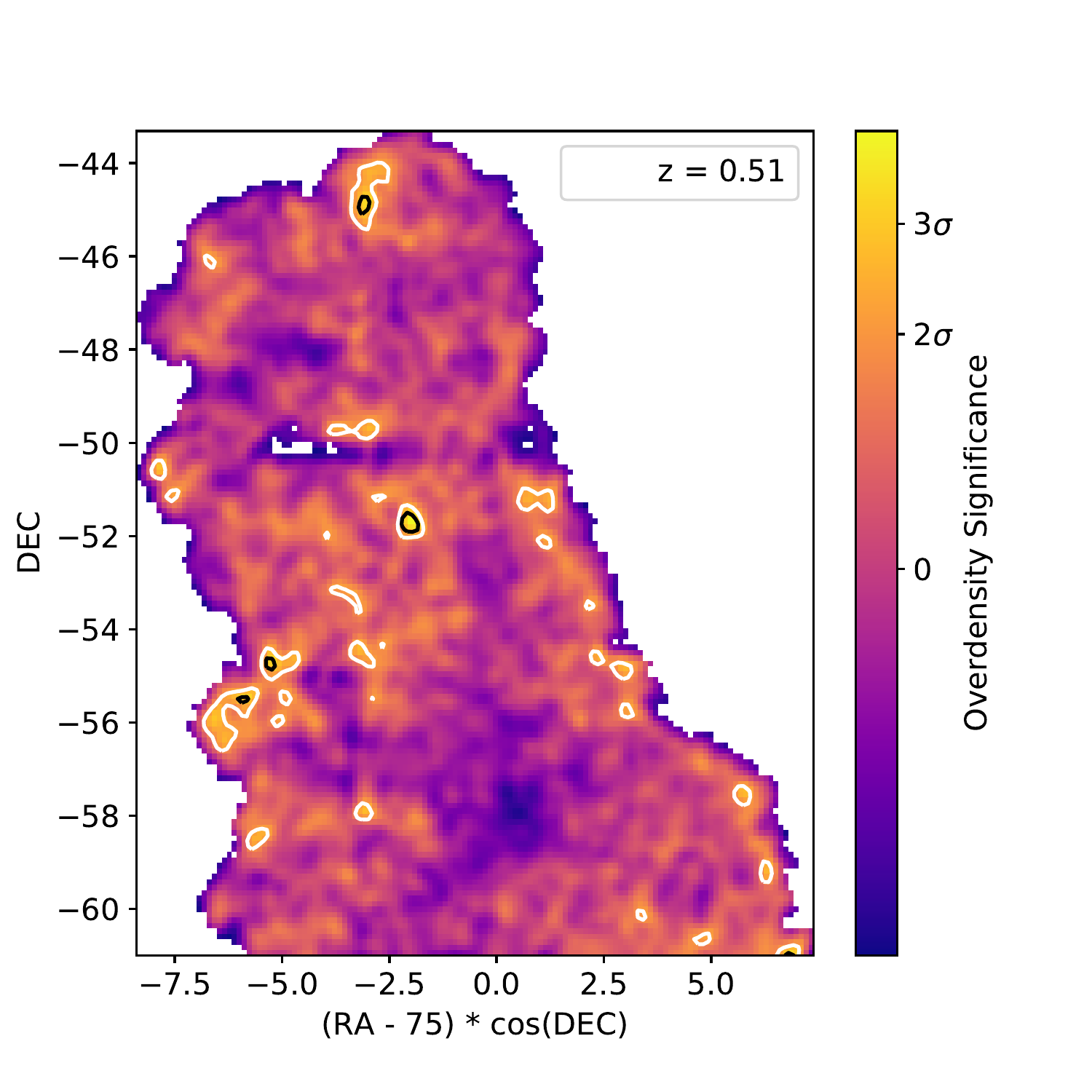}} \\
\subfigure[][$z = 0.50 - 0.70$]{%
\label{fig:5070}%
\includegraphics[width=2in]{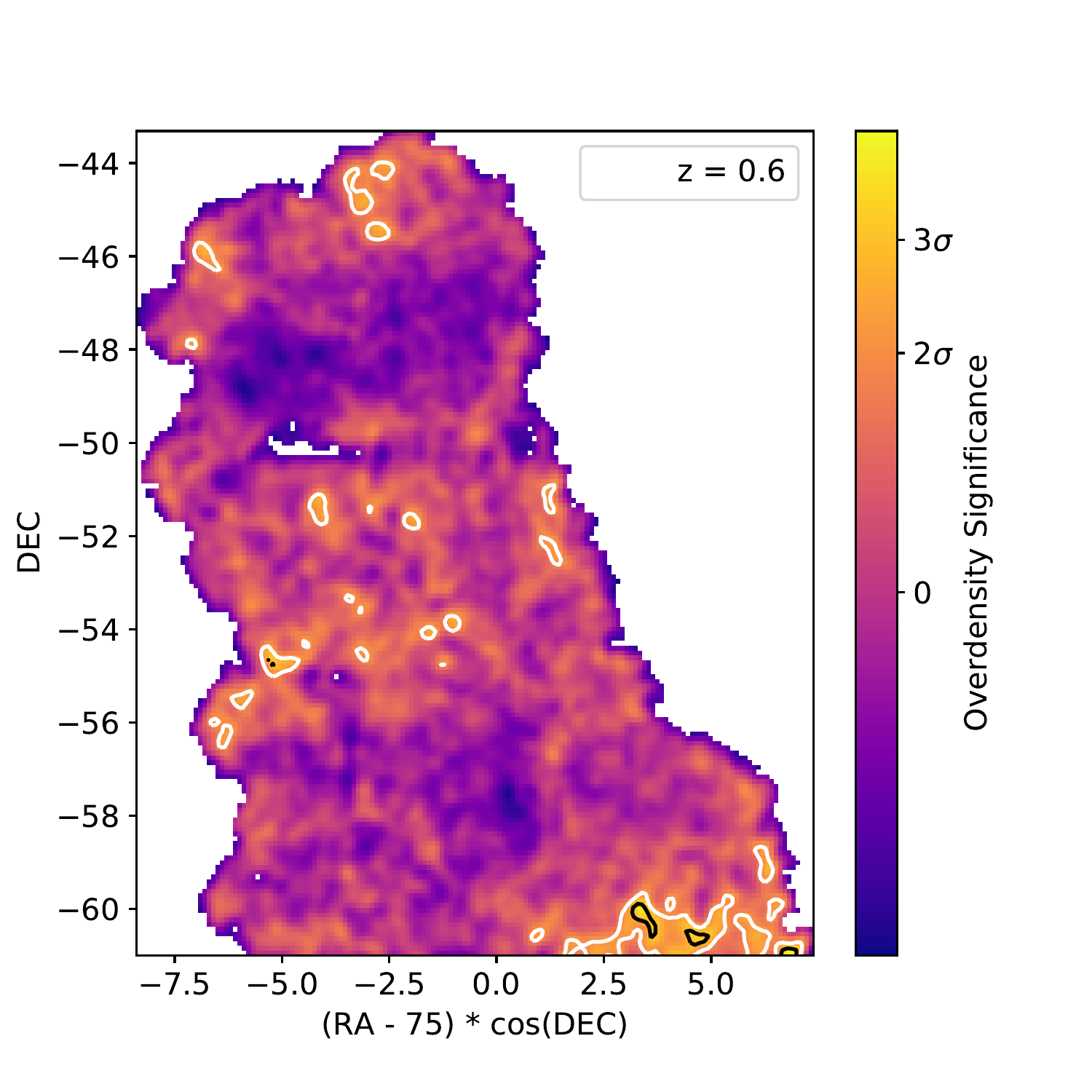}}%
\hspace{8pt}%
\subfigure[][$z = 0.59 - 0.79$]{%
\label{fig:5979}%
\includegraphics[width=2in]{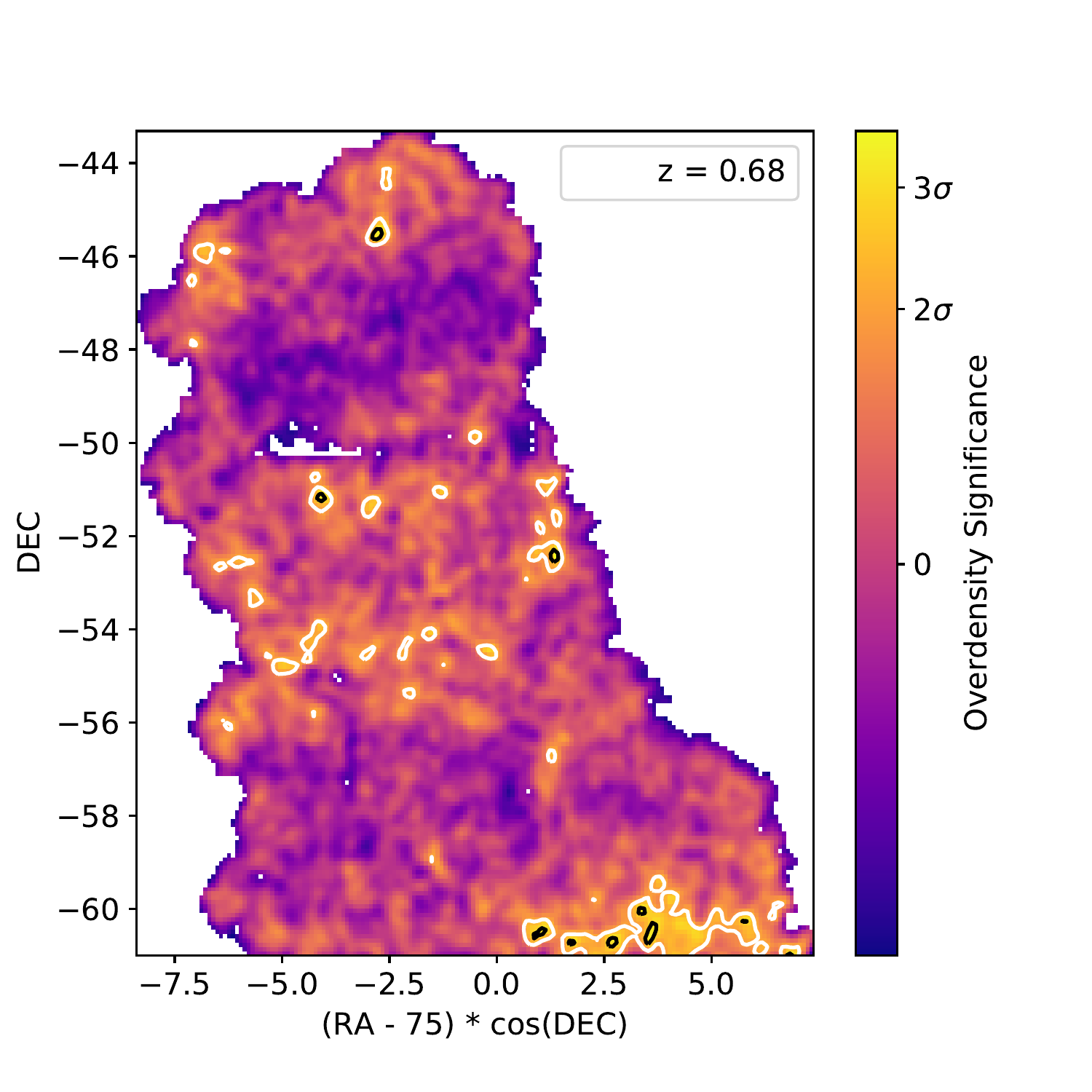}}%
\hspace{8pt}%
\subfigure[][$z = 0.71 - 0.91$]{%
\label{fig:7191}%
\includegraphics[width=2in]{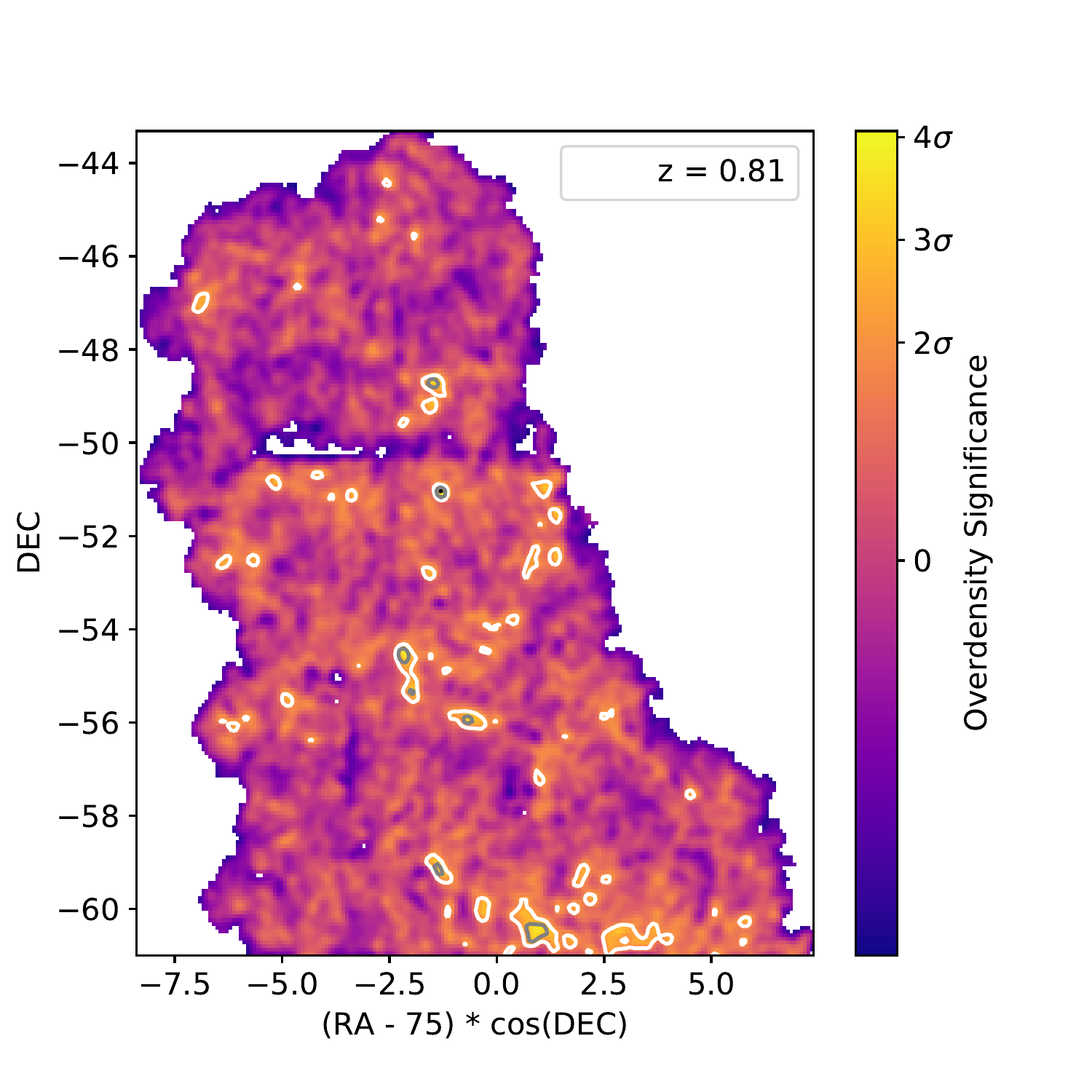}} \\
\subfigure[][$z = 0.80 - 1.00$]{%
\label{fig:80100}%
\includegraphics[width=2in]{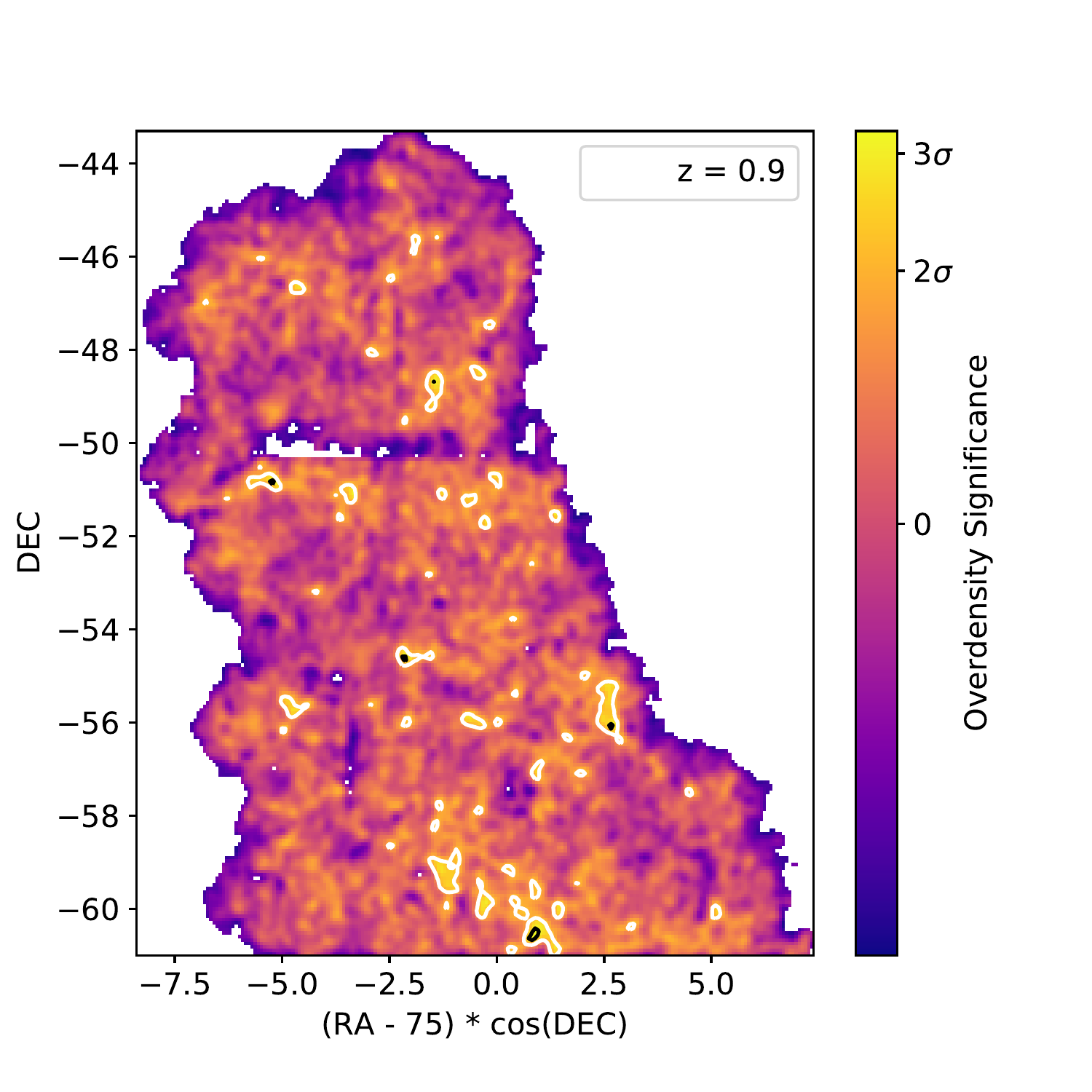}}%
\hspace{8pt}%
\subfigure[][$z = 0.89 - 1.09$]{%
\label{fig:89109}%
\includegraphics[width=2in]{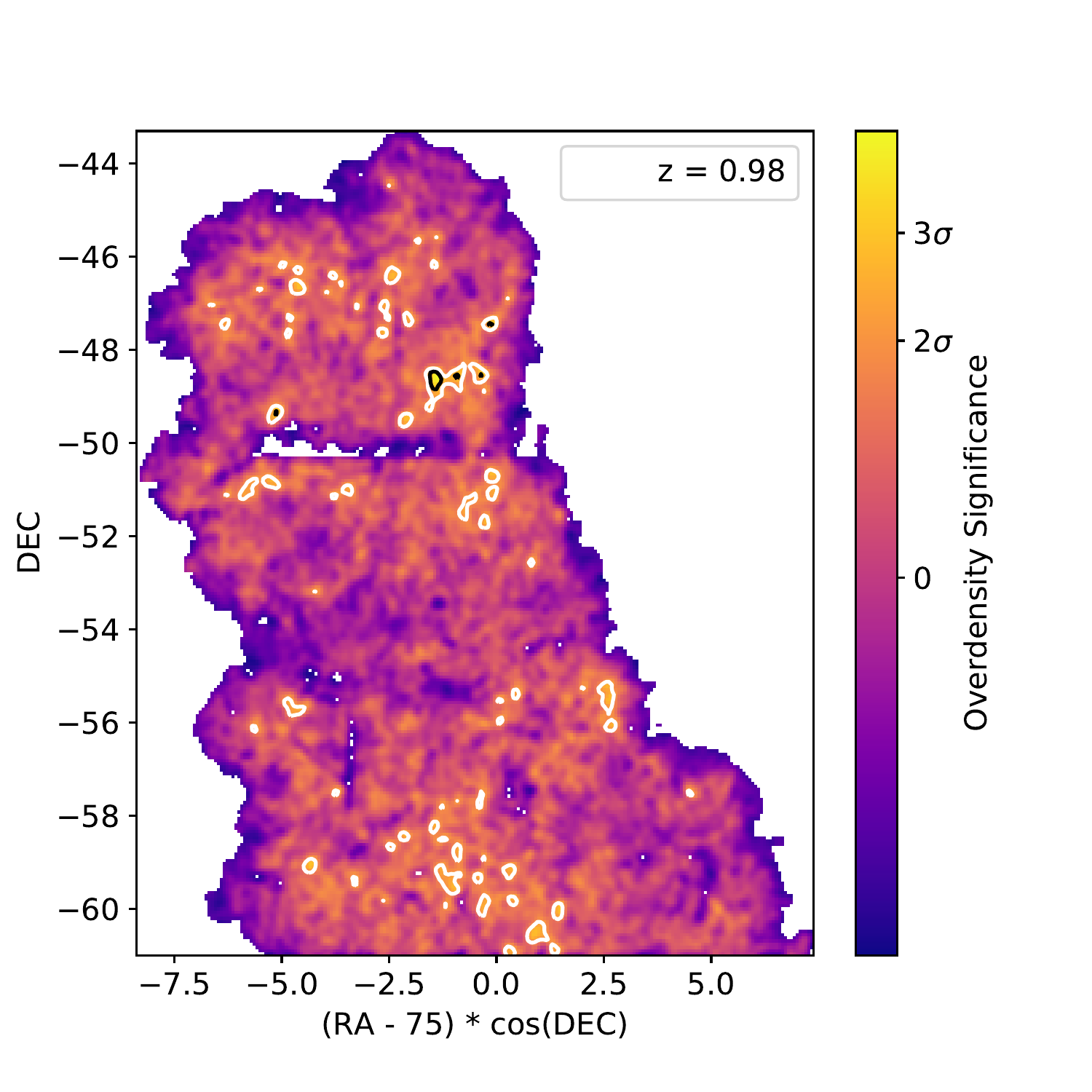}}%
\hspace{8pt}%
\subfigure[][$z = 1.01 - 1.21$]{%
\label{fig:101121}%
\includegraphics[width=2in]{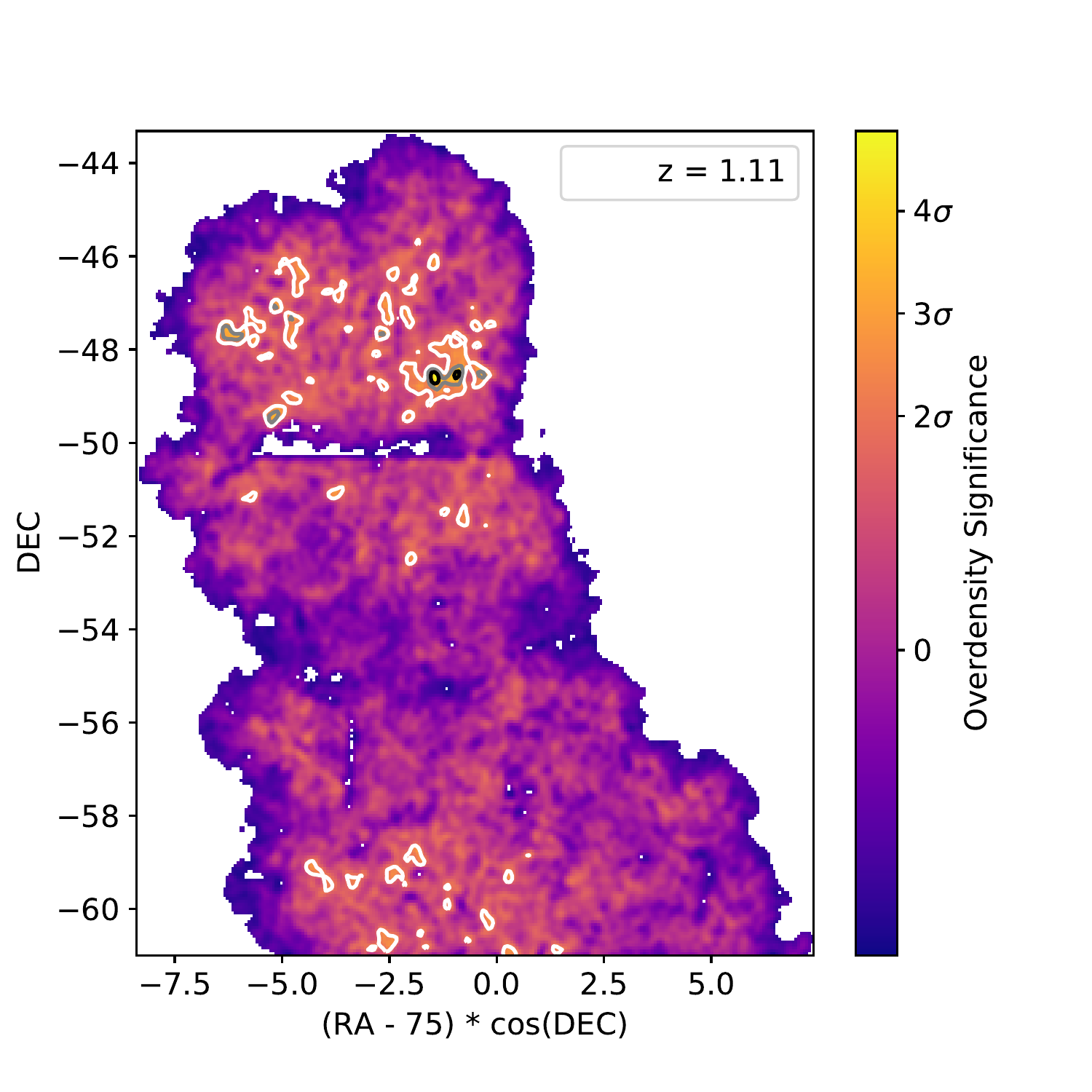}}%
\caption{Galaxy density significance maps at each redshift slice. The legend in each figure is the mean value of the redshift of the galaxies in the galaxy bins. Each figure is made in a redshift slice of 0.2 width. We can see that the same structure might be detected in multiple redshift slices. The contours represent two,three and four $\sigma$ overdensity levels.}
\label{fig:all_contour}
\end{figure*}

Here we present the overdensity contour plots from redshift slices between $z=0.2$ and $z=1.2$ (Fig. \ref{fig:all_contour}). We note another supercluster candidate between redshift $0.5 and 0.8$.

\end{document}